\documentclass[default]{sn-jnl}

\usepackage{amsmath}

\jyear{2024}%

\theoremstyle{thmstyleone}%
\theoremstyle{thmstyletwo}%
\newcommand{\rsun}{$R_{\odot}$}
\usepackage{hyperref}
\usepackage{amsmath}
\usepackage{amstext}
\usepackage{graphicx}
\usepackage{xcolor}
\usepackage{comment}
\usepackage{caption}
\usepackage{subcaption}
\usepackage{makecell}
\usepackage{mathtools}
\theoremstyle{thmstylethree}%
\usepackage{natbib}

\begin{document}

\title[Calibration of Spectropolarimetry channel of VELC onboard Aditya-L1]{Calibration of Spectropolarimetry channel of Visible Emission Line Coronagraph onboard Aditya-L1}

\author[1]{\fnm{Venkata Suresh} \sur{Narra}}

\author*[1]{\fnm{K.} \sur{Sasikumar Raja}}\email{sasikumar.raja@iiap.res.in}
\author[1]{\fnm{Raghavendra Prasad} \sur{B}}
\author[1]{\fnm{Jagdev} \sur{Singh}}
\author[1]{\fnm{Shalabh} \sur{Mishra}}
\author[1]{\fnm{Sanal Krishnan} \sur{V U}}
\author[1]{\fnm{Bhavana Hegde} \sur{S}}
\author[1]{\fnm{Utkarsha} \sur{D.}}
\author[1]{\fnm{Natarajan} \sur{V}}
\author[1]{\fnm{Pawan Kumar} \sur{S}}
\author[1]{\fnm{Muthu Priyal} \sur{V}}
\author[1]{\fnm{Savarimuthu} \sur{P}}
\author[2]{\fnm{Priya} \sur{Gavshinde}}
\author[1]{\fnm{Umesh Kamath} \sur{P}}

\affil*[1]{\orgname{Indian Institute of Astrophysics}, \orgaddress{\street{2nd Block, Koramangala}, \city{Bangalore}, \postcode{560034}, \state{Karnataka}, \country{India}}}


\abstract{The magnetic field strength and its topology play an important role in understanding the formation, evolution, and dynamics of the solar corona. Also, it plays a significant role in addressing long-standing mysteries such as coronal heating problem, origin and propagation of coronal mass ejections, drivers of space weather, origin and acceleration of solar wind, and so on. Despite having photospheric magnetograms for decades, we do not have reliable observations of coronal magnetic field strengths today. To measure the coronal magnetic field precisely, the spectropolarimetry channel of the Visible Emission Line Coronagraph (VELC) on board the Aditya-L1 mission is designed. Using the observations of coronal emission line Fe XIII [10747{\AA~}], it is possible to generate full Stokes maps (I, Q, U, and V) that help in estimating the Line-of-Sight (LOS) magnetic field strength and to derive the magnetic field topology maps of solar corona in the Field of View (FOV) (1.05 -- 1.5~R$_{\odot}$). In this article, we summarize the instrumental details of the spectropolarimetry channel and detailed calibration procedures adopted to derive the modulation and demodulation matrices. Furthermore, we have applied the derived demodulation matrices to the observed data in the laboratory and studied their performance.
}




\maketitle

\section{Introduction}

Aditya-L1 is the first Indian mission to explore the Sun and solar atmosphere and was launched on 2 September 2023 from the Sriharikota launch pad, India. It has seven payloads onboard that cover electromagnetic spectra ranging from hard X-rays to infrared frequencies \citep{Seetha2017, BRP2017, Durgesh2022}. Four of the seven payloads are remote sensing instruments, and the others are in-site payloads. Aditya-L1 monitors the Sun and the solar atmosphere from the first Lagrangian point (L1), which is 1.5 million kilometers from Earth, enabling us to observe uninterruptedly. The Visible Emission Line Coronagraph is an internally occulted reflective coronagraph. VELC has four channels: (i) one imaging/continuum channel to image the solar corona at 5000\AA~~ in the Field-Of-View (FOV) 1.05-3 \rsun, (ii) three spectroscopy channels at 5303\AA~, 7892\AA~ and 10747\AA~ in FOV 1.05-1.5 \rsun, and (iii) one spectropolarimetry channel (VELC/SP) at 10747\AA~ in FOV 1.05-1.5 \rsun.

In this article, our primary focus is on the spectropolarimetry channel of VELC. The main objectives of this channel are to estimate the magnetic field strength and to understand the magnetic topology of the solar corona. Despite having the photospheric magnetogram data for decades, our understanding of the coronal magnetic fields remains limited. It is the first time space-based spectropolarimetry observations that are planned to measure the coronal magnetic fields. Earlier, \citep{Lin2000, Lin2004} reported coronal magnetic fields using ground-based 0.46 off-axis reflecting coronagraph data. A comprehensive review of coronal magnetic fields was also provided by Sasikumar Raja et al 2022 \cite{Sas2022}.

\section{VELC and Spectropolarimetry Channel}

\begin{figure}[!ht]
\centering
   \begin{subfigure}[b]{1.0\textwidth}
   \includegraphics[width=1.0\textwidth]{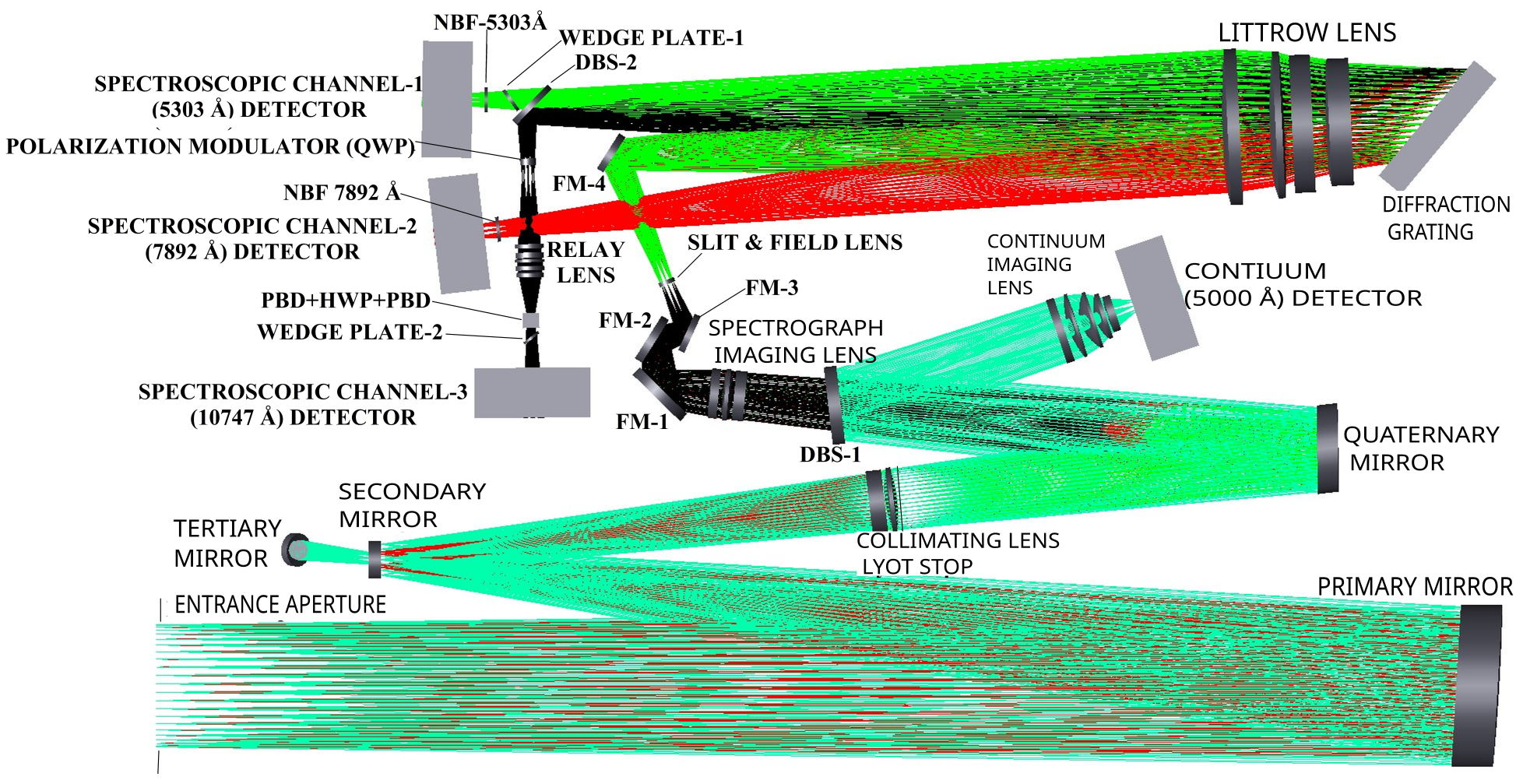}
   \caption{}
   \label{fig:optical_layout}
\end{subfigure}

\begin{subfigure}[b]{1.0\textwidth}
   \includegraphics[width=1.0\textwidth]{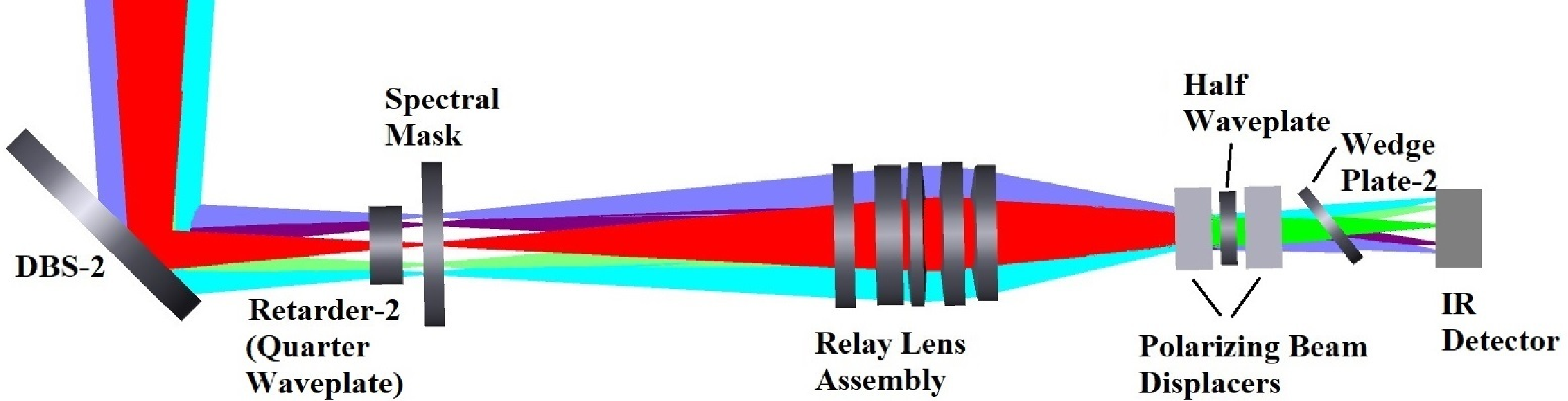}
   \caption{}
   \label{fig:sp_layout}
\end{subfigure}
\caption{(a) Optical layout the VELC. (b) Optical layout of the spectropolarimetry channel.}
\end{figure}

The optical layout of VELC \citep{Rajkumar2018, Singh2019} and the optical layout of the spectropolarimetry channel \citep{Venkata2022} are shown in Figures \ref{fig:optical_layout} and \ref{fig:sp_layout}. The light from the Sun and solar corona that fall on the primary mirror reimages on to the M2 in which there is an elliptical hole that exits the disk light and reflects the coronal light on to the Quaternary Mirror (QM). The light reflected from QM is imaged on to the continuum channel detector. A part of the light is reimaged on to slits by the spectrograph imaging lens assembly via three fold mirrors (FM1, FM2 and FM3). The spectra from the grating is imaged on to the three detectors. The IR light at 10747 \AA~ reflected by the Dichroic Beam Splitter-2 feed to the spectropolarimetry channel of VELC. The reflected light is transmitted through a retarder (Quarter Wave Plate) which is mounted on a rotating mechanism. The QWP rotates at a speed of 5.95 RPM. Since VELC has four slits, in order to minimize the spectral overlapping from the adjacent ones, a spectral mask is kept after the retarder which then followed by a relay lens assembly to re-image the beam onto the IR detector through a combination of PBD – HWP – PBD, where, PBD is Polarizing Beam Displacer and Half Wave Plate (HWP). Note that PBD – HWP – PBD combination acts as an analyzer. Each PBD provides the required separation between ordinary (O) and extra-ordinary (E) rays. The HWP flips the polarization states of the output beams from PBD-1 and ensures that there is no path difference between the O and E rays at the final focal plane. A wedge plate is used to reduce the aberrations produced by the PBD-HWP-PBD combination. The InGaAs IR detector with $512 \times 640$ pixels is placed in the image plane. The size of the pixels is $25 \mu m \times 25 \mu m$ with a spectral range of 9000 \AA~ to 1700 \AA~. The detector operates in rolling shutter mode and is capable of providing images in both low and high gain modes \citep{Shalabh2024}.

The spectroscopy and spectropolarimetry channels can be operated in the Sit-and-Stare (SS) mode and Raster Scan (RS) modes. In the SS mode the Linear Scan Mechanism (LSM) moves to a desired location so the image onto the slits moves. The spectra from the desired location will be imaged on the detectors. In the case of the RS mode, the LSM moves from +1mm to -1mm in multiple steps of $10 \mu m$ with minimum step size $10 \mu m$. At every step, it wails till all detector acquires the given number of images with a desired exposure time. RS mode of observations are useful to make the raster scan images.

\section{Experimental Setup and Method}

\begin{figure}[!ht]
    \centering
    \includegraphics[width=0.8\textwidth]{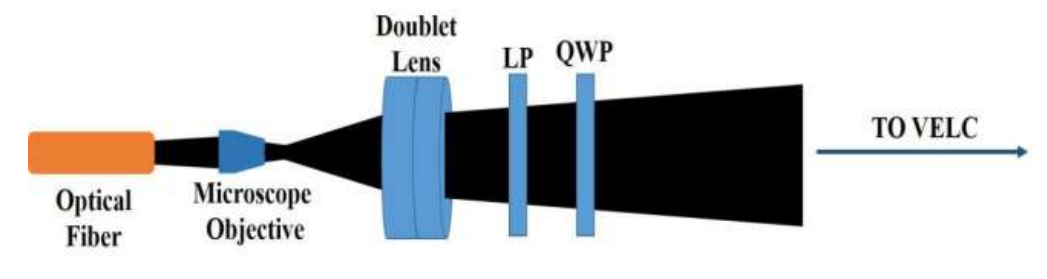}
    \caption{The experimental setup used to measure the instrument response matrix, modulation and demodulation matrices is shown.}
    \label{fig:setup}
\end{figure}

First, we have set up a Polarization State Generator (PSG) consisting of one fixed linear polarizer and one rotatable quarter-wave plate to generate different polarization states. The polarized light generated by PSG for different angles of QWP is fed to the VELC as shown in \ref{fig:setup}. At every polarization state of PSG, we rotated the VELC QWP, whose angular speed is 5.95 rpm, and acquired the data. Exposure times are suitably selected to combine the number of frames acquired in a single sector that remains within the angle of $22.5^\circ$. Note that there are 16 hall sensors that sense the angle information during the retarder movement.

Note that VELC retarder makes one rotation in 10084 ms. Therefore, we selected the exposure time of 203 ms. In one full rotation, it acquires approximately 98 frames. Therefore, we get six frames per sector (an angular span of $22.5^\circ$). Also, we acquired the dark data before and after taking the spectropolarimetry data.

\section{Results and Discussions}\label{sec4}

To estimate the modulation, demodulation, and instrument response matrices and the spectropolarimetric efficiencies, we acquired the data during vacuum calibration of VELC at Prof. MGKM Lab, CREST campus of the Indian Institute of Astrophysics. During this test, the detector chip was cooled to $-17^\circ$.

\begin{figure*}[!ht]

   \begin{subfigure}[b]{\textwidth}
   \centering
   \includegraphics[width=0.6\textwidth]{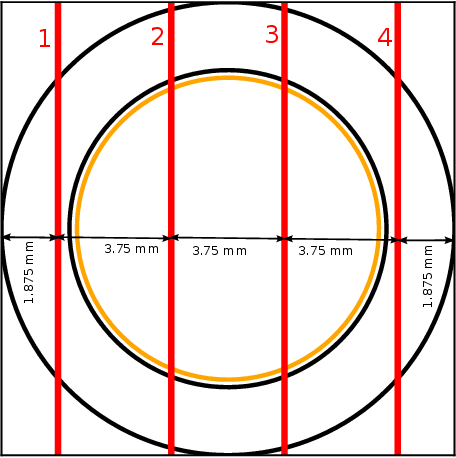}
   \caption{}
   \label{fig:slits}
\end{subfigure}
\begin{subfigure}[b]{\textwidth}
   \includegraphics[width=1.1\textwidth]{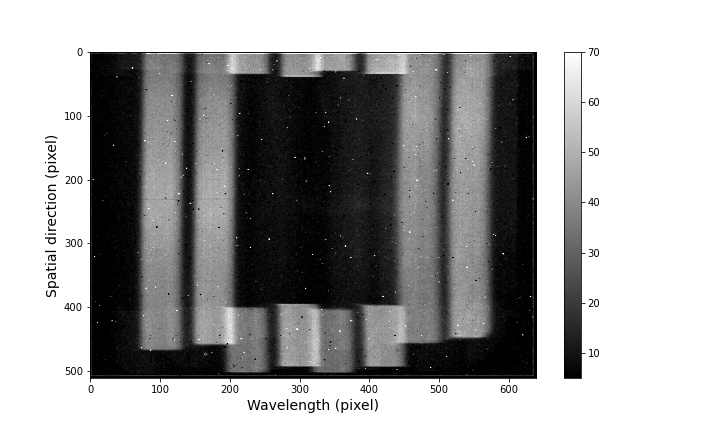}
   \caption{}
   \label{fig:spimage}
\end{subfigure}
\caption{(a) The four-slit configuration of VELC is shown. The yellow circle is the photosphere, and two concentric black circles are 1.05 and 1.5 \rsun. The red vertical lines are the slits. (b) The spectra obtained in the laboratory are shown. The eight spectra are the E and O ray spectra of four slits.}
\end{figure*}

\subsection{Generation of Input Polarization States}
In order generate different polarization states we designed a Polarization State Generator (PSG) setup. It includes one linear polarizer and one rotatable QWP. Knowing the theoretical Mueller matrices for the linear polarizer ($M_{LP}$) and QWP ($M_{QWP}$), we have obtained a generalized Stokes vector ($S_{in}$) of the setup using equation \ref{eq:mm}.

Note that we have used a Stokes vector for unpolarized light, i.e., $[1~ 0~ 0~ 0]^T$ in the calculation.
\begin{equation}\label{eq:mm}
   S_{in} =  M_{QWP}~ M_{LP} ~[1~ 0~ 0~ 0]^T
\end{equation}
where,

\begin{equation}\label{eq:qwp}
M_{QWP}=  \begin{bmatrix}
1&0&0&0\\
0& \cos^2(2\theta)& \sin(2\theta)\cos(2\theta) & \sin(2\theta)\\
0& \sin(2\theta)\cos(2\theta) & \sin^2(2\theta)& -\cos(2\theta)\\
0& -\sin(2\theta)& \cos(2\theta)& 0 \\
\end{bmatrix}
\end{equation}

and

\begin{equation}\label{eq:lp}
M_{LP}= \begin{bmatrix}
1&1&0&0\\
1&1&0&0\\
0&0&0&0\\
0&0&0&0\\
\end{bmatrix}
\end{equation}

Note that $\theta$ is the polarization angle, and $T$ is the transpose of the matrix. The derived generalized solution of Equation \ref{eq:mm} is,
\begin{equation}\label{eq:gsin}
S_{in}= \begin{bmatrix}
1\\
\cos^2(2\theta)\\
\sin(2\theta)\cos(2\theta)\\
-\sin(2\theta)\\
\end{bmatrix}
\end{equation}

Knowing $S_{in}$, we obtained the Stokes vectors for different $\theta$ ranging from $0^\circ$ to $180^\circ$ with a step size of $15^\circ$ and are shown below.

\begin{equation}\label{eq:sin}
S_{in} =
\begin{bmatrix}
1 & 1 & 0 & 0 \\
1 & 0.883 & 0.3214 & -0.342 \\
1 & 0.5868 & 0.4924 & -0.6428 \\
1 & 0.25 & 0.433 & -0.866 \\
1 & 0.0302 & 0.171 & -0.9848 \\
1 & 0.0302 & -0.171 & -0.9848 \\
1 & 0.25 & -0.433 & -0.866 \\
1 & 0.5868 & -0.4924 & -0.6428 \\
1 & 0.883 & -0.3214 & -0.342 \\
1 & 1 & 0 & 0 \\
1 & 0.883 & 0.3214 & 0.342 \\
1 & 0.5868 & 0.4924 & 0.6428 \\
1 & 0.25 & 0.433 & 0.866 \\
1 & 0.0302 & 0.171 & 0.9848 \\
1 & 0.0302 & -0.171 & 0.9848 \\
1 & 0.25 & -0.433 & 0.866 \\
1 & 0.5868 & -0.4924 & 0.6428 \\
1 & 0.883 & -0.3214 & 0.342 \\
\end{bmatrix}
\end{equation}

\subsection{Modulation Matrix}

The light with known polarization states that is listed in the above section was illuminated the Entrance Aperture of the VELC and acquired the data in the IR channel with an exposure time of 203ms. Upon acquiring the data we have initiated the movement of the retarder. Since it takes 10084 ms to complete one rotation, the detector can acquire approximately 50 frames per rotation.  Therefore, we get six images per sector (an angular space of $22.5^\circ$). VELC is designed to have four slits such that it observes simultaneously at four spatial locations of the corona as shown in Figure \ref{fig:slits}. Furthermore, VELC/SP is a dual beam spectrometer and therefore splits four spectra into O-ray and E-ray. Therefore, in a single image, we get eight spectra as shown in Figure \ref{fig:spimage}. After acquiring the light images, we have performed the dark subtraction. For this, 16 dark images with 203ms exposure time was acquired. Using the dark images, the master dark image was created by averaging all images and subtracted it from the light images.

We selected a region of E-ray spectra (i.e., 310 pixels) in the spatial direction (Y-axis) and 45 pixels along the spectral direction (X-axis), and found mean counts ($I_e$) for all the images. Similarly,  mean counts of the O-ray spectra ($I_o$) over the pixels specified above are measured. These intensities were further averaged over each sector of $22.5^\circ$ and generated a $I_{mod}$ matrix separately for both O ray and E ray.


Knowing the $I_{mod}$ and $S_{in}$ matrices, we have derived the optimum modulation matrices for both O and E modes separately.

\begin{equation}
    O=I_{mod} S_{in}^T (S_{in}S_{in}^T)^{-1}
\end{equation}

The normalized modulation matrices derived using O-mode ($O_O$) observations,

\begin{equation}\label{eq:sin}
\text{O}_O =
\begin{pmatrix}
1 & 0.51472348 & 0.16614091 & 0.22969513 \\
1 & -0.43329999 & 0.38728775 & 0.77480626 \\
1 & -0.35876231 & 0.70062804 & 0.20589564 \\
1 & 0.51323429 & 0.36768651 & -0.1396031 \\
1 & 0.60468295 & -0.19695827 & -0.11201827 \\
1 & -0.09619814 & -1.11688079 & -0.44184011 \\
1 & 0.04239088 & -0.60683106 & -0.89084785 \\
1 & 0.66869627 & 0.04797316 & -0.35921256 \\
\end{pmatrix}
\end{equation}

The normalized modulation matrices derived using E-mode ($O_E$) observations,

\begin{equation}\label{eq:sin}
\text{O}_E =
\begin{pmatrix}
1 & 0.17688466 & -0.16009568 & -0.91119486 \\
1 & 1.04152682 & -0.17497231 & -0.72680567 \\
1 & 1.07359391 & -0.4698261 & -0.31301888 \\
1 & 0.17008938 & -0.85011471 & 0.14208459 \\
1 & 0.06855784 & 0.83111976 & -0.00135888 \\
1 & 0.74781198 & 0.83840855 & 0.1507165 \\
1 & 0.67812993 & 0.59968176 & 0.51842439 \\
1 & -0.19253052 & 0.25263437 & 0.71490065 \\
\end{pmatrix}
\end{equation}

\subsection{Demodulation Matrix}
After measuring the normalized modulation indices (i.e., $O_O$ and $O_E$) we measured the demodulation matrices (D) using ,

\begin{equation}
    D=(O^T O)^{-1}O^T
\end{equation}

The mean of de-modulation matrix $D_O$
\begin{equation}\label{eq:sin}
\text{D}_O=
\begin{pmatrix}
0.0886 & 0.3844 & -0.1494 & 0.4174 \\
0.2142 & -0.243 & -0.162 & 0.5463 \\
0.1952 & -0.4985 & 0.5129 & -0.3994 \\
0.0825 & 0.1759 & 0.2982 & -0.217 \\
0.0727 & 0.3926 & -0.2182 & 0.2837 \\
0.1559 & -0.1522 & -0.5886 & 0.2364 \\
0.1305 & -0.3436 & 0.1474 & -0.6724 \\
0.0604 & 0.2846 & 0.1596 & -0.1951 \\
\end{pmatrix}
\end{equation}

The mean of de-modulation matrix $D_E$ \\
\begin{equation}\label{eq:sin}
\text{D}_E=
\begin{pmatrix}
0.2849 & -0.4129 & 0.0554 & -0.5324 \\
-0.0100 & 0.2643 & -0.0068 & -0.2132 \\
-0.0315 & 0.3932 & -0.2148 & 0.0990 \\
0.2565 & -0.1509 & -0.4456 & 0.2296 \\
0.2209 & -0.2989 & 0.3126 & -0.2034 \\
-0.0120 & 0.2344 & 0.2705 & 0.0481 \\
0.0001 & 0.2765 & 0.0982 & 0.2976 \\
0.2911 & -0.3059 & -0.0694 & 0.2747 \\
\end{pmatrix}
\end{equation}

\subsection{Evaluation of the measured demodulation matrices}

\begin{figure}[!ht]
    \centering
    \includegraphics[width=1\textwidth]{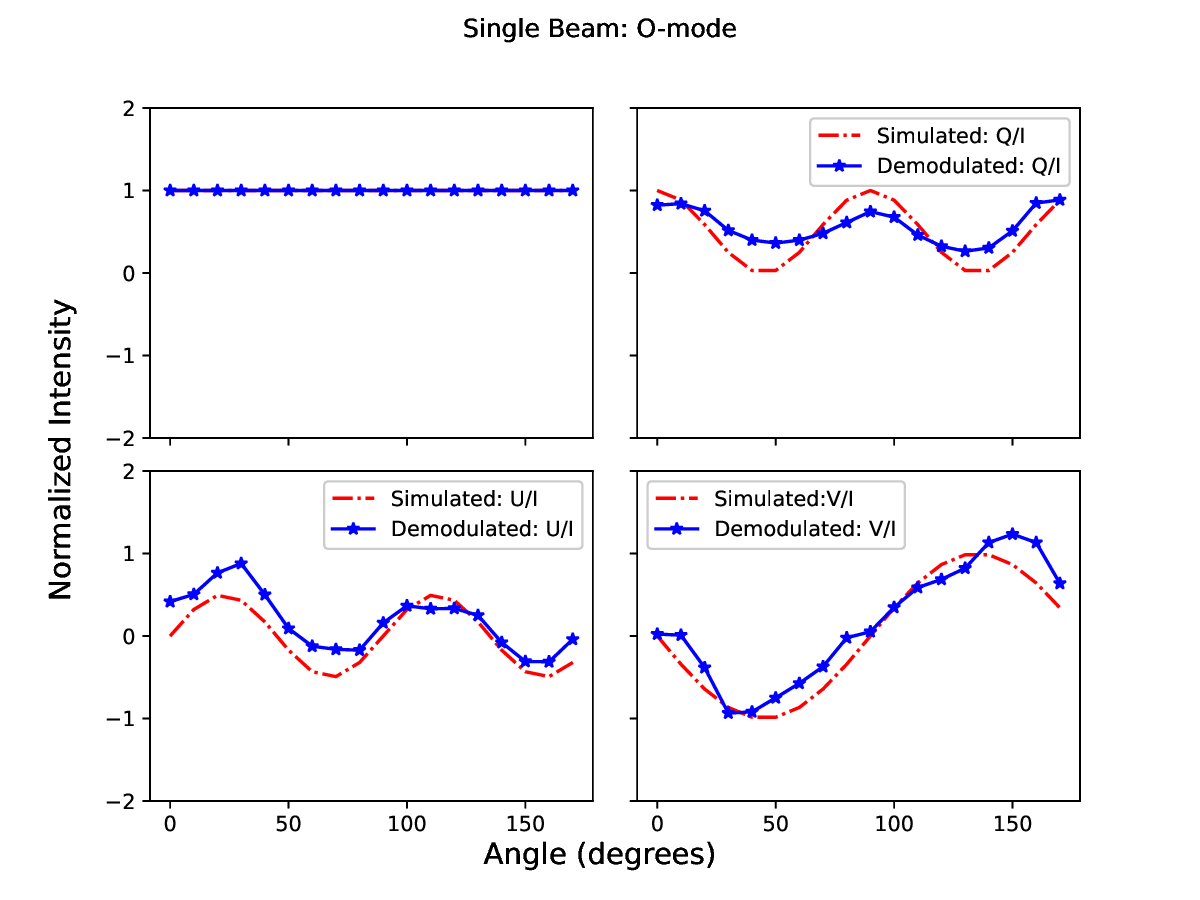}
    \caption{Four subplots show the Stokes I/I, Q/I. U/I and V/I parameters derived using the O mode observations. The curve in red color indicate the input Stokes parameters and the blue color curves indicate the demodulated Stokes parameters.}
    \label{fig:omode}
\end{figure}

\begin{figure}[!ht]
    \centering
    \includegraphics[width=1\textwidth]{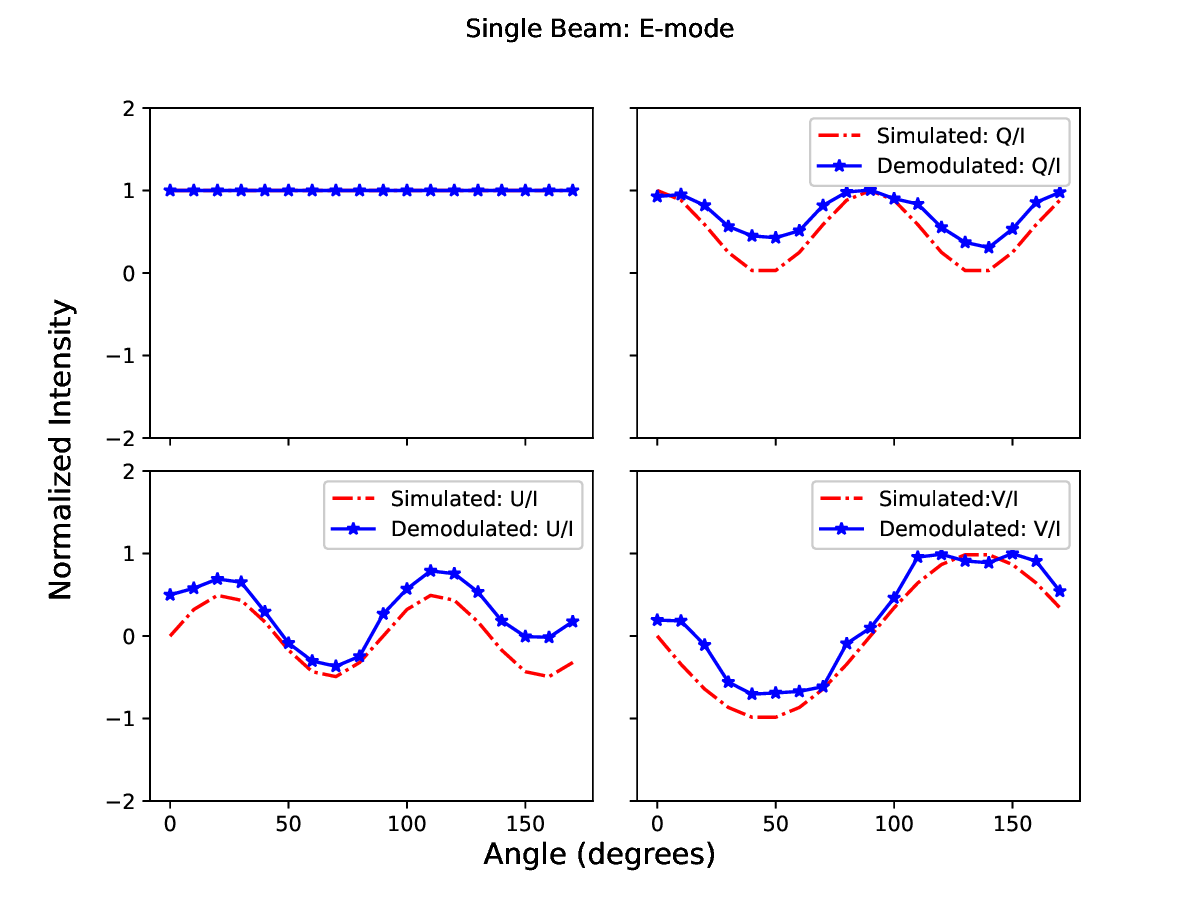}
    \caption{Four subplots show the Stokes I/I, Q/I. U/I and V/I parameters derived using the E mode observations. The curve in red color indicate the input Stokes parameters and the blue color curves indicate the demodulated Stokes parameters.}
    \label{fig:emode}
\end{figure}

To verify the above results, we have multiplied the observations carried out from the Prof. MGKM Laboratory with the demodulation matrices and obtained the Stokes vectors that are shown in Figures \ref{fig:omode} and \ref{fig:emode}. The four subplots in the Figures are normalized Stokes I, Stokes Q, Stokes U and Stokes V profiles. The red-colored curves show the theoretically derived Stokes inputs and blue-colored curves show the demodulated Stokes parameters.

\section{Summary and Conclusions}

The Visible Emission Line Coronagraph on board Aditya-L1 has a dedicated spectropolarimetry channel at 10747 \AA~~. The primary goal of this channel is to make full Stokes maps of the solar corona in FOV 1.05-1.5 \rsun. Such Stokes maps are crucial in estimating the coronal magnetic fields and topology. In this article, we have summarized the spectropolarimetry channel of VELC and the experimental setup used to calibrate and to measure the demodulation matrix of the instrument. We acquired the data by feeding the different input polarization states by using a linear polarizer and rotating the QWP combination. For every input state of polarization, the data were acquired by rotating the QWP of VELC. Furthermore, we have summarized the way modulation, and demodulation matrices separately for O-mode and E-mode is obtained. In addition, we have evaluated the measured demodulation matrix by applying them to the observations.

\noindent {\bf Acknowledgments:}
We thank all the Scientists/Engineers at the various centres of
ISRO such as URSC, SAC, LEOS, VSSC etc. and Indian Institute of Astrophysics who have made great contributions to the mission in order to achieve the present state. Authors acknowledge Drs. K. Nagaraju, C. Kathiravan, Jayant Joshi, and K. Sankarasubramanian for their valuable suggestions. \\

\noindent {\bf Author's contributions:} NVS and KSR designed the setup, analyzed the data, and prepared the manuscript. BRP and JS provided valuable suggestions on the test setup and calibration technique and reviewed the manuscript. SM, SKVU, KSR, BHS, UD and SP helped with the setup and operated the detectors and retarder mechanism to acquire the data at Prof. MGK Menon Laboratory, CREST campus, Indian Institute of Astrophysics. NV, PKS, UKP helped with the mechanical setup for the experiment. MPV and PG reviewed the manuscript.\\ 

\noindent {\bf Funding} We appreciate the financial support from the Indian Space Research Organization (ISRO), India for this project.

\section*{Declarations}

{\bf Competing interests:} The authors declare no competing interests.\\ \\
{\bf Conficts of interest:} The authors declare that they have no confict of interest

\bibliographystyle{unsrt}
\bibliography{main.bib}

\end{document}